# The Dynamics-Based Approach to Studying Terrestrial Exoplanets

submitted to the Exoplanet Task Force (AAAC), 2 April 2007

David Charbonneau[1] & Drake Deming[2]

## Motivation

The study of planets orbiting nearby, main-sequence stars has proceeded at a breakneck pace: Less than ten years elapsed from the first discovery of a sub-Jupiter-mass companion to a Sun-like star (Mayor & Queloz 1995) to the first direct detection of light emitted from such an exoplanet (Deming et al. 2005; Charbonneau et al. 2005). This field of astrophysics is currently in an observationally-driven phase, and the bulk of research activity in the past decade has been fueled initially by precise radial-velocity measurements (Udry, Fischer, & Queloz 2007), and, more recently, by the combination of that technique with transit photometry (Charbonneau et al. 2007). Planetary systems for which both the radial-velocity orbit and the transit light curve are measured permit us to determine the mass and radius of the planet, which in turn yield powerful constraints on its physical structure and bulk composition. The true power of a transiting geometry, however, is that it permits the study of the planetary atmosphere without the need to spatially isolate the light from the planet from that of the star. It is this technique, which we term "occultation spectroscopy", that yielded the first emergent spectra of planets orbiting nearby Sun-like stars (Richardson et al. 2007; Grillmair et al. 2007; Swain et al. 2007).

The unifying feature of all the exoplanet techniques that have borne fruit to date for mature, Gyr-old stars is that the experimental design is based on measuring temporal changes to intrinsic quantities that arise due to the planetary orbit. We unite these techniques under the broad terminology of "dynamics-based methods", and contrast them with another broad class of methods, namely "imaging techniques" (e.g. Beuzit et al. 2007), which include adaptive optics, coronography, and certain applications of interferometry. One of the great quests of astronomy is to obtain the spectrum of a terrestrial planet orbiting within the habitable zone of its star, and the dominant challenge in doing so is to isolate the light of the planet from that of the star. Dynamics-based methods separate these signals temporally, whereas imaging techniques do so spatially. In light of the overwhelming dominance of dynamics-based methods over the past decade, we challenge the notion that spectra of terrestrial planets necessarily require extreme imaging methods. We advocate that some resources be committed to refining the proven technologies of radial-velocity measurements, transit photometry, and occultation spectroscopy (i.e. emergent infrared spectra obtained at secondary eclipse). We believe that it is these methods, not imaging, that will yield the first detections of terrestrial exoplanets orbiting within their stellar habitable zones, as well as the first observations of the spectra of such planets.

We see four broad advantages of the dynamics-based approach over imaging techniques:

1. By obviating the requirement for extreme contrast ratio at very small angular separations, it is technologically much simpler and hence significantly less expensive.
2. Much of the science may be accomplished with facilities that are either in operation, or general purpose observatories that are in advanced stages of planning or construction.
3. Occultation spectroscopy permits a direct estimate of the planetary surface flux with no degeneracy between temperature and emitting area.

---

1*Dept of Astronomy, Harvard Univ, 60 Garden St, Cambridge, MA 02138; dcharbon@cfa.harvard.edu*
2*NASA Goddard Space Flight Center, Greenbelt, MD 20771; drake.deming@gsfc.nasa.gov*

4. Precise estimates of the masses and radii of planets studied by occultation spectroscopy will be determined from radial velocities and transit observations. Given these detailed constraints on the physical structure and bulk composition, the inferences about the atmosphere from the observed spectra are likely to be far more penetrating than that for cases in which only the spectrum is available.

**The M-dwarf Opportunity**

We see a particularly attractive opportunity in M-dwarfs. Such stars are by far the most common in the local solar neighborhood. The most recent results from the RECONS Survey (Henry et al. 2007) report 348 stars within 10 pc (as determined from trigonometric parallaxes), of which 239 are M dwarfs and only 21 are G dwarfs. Projecting these numbers by volume, we expect 10,000 M-dwarf stars within 35 pc. This estimate is consistent with the number of M-dwarfs in that volume identified by large proper motions and 2MASS photometry (Lepine & Shara 2005; Lepine 2005) but for which parallaxes have not yet been obtained. Whether these low-mass stars have the same rate-of-occurrence of planetary companions as Sun-like stars is an open question. Some authors (Butler et al. 2004) have stated that they find the rate of Jupiter-mass planets to be significantly suppressed for M-dwarf primaries, but others state clearly that the survey results do not yet permit this conclusion (Endl et al. 2006; Bonfils et al. 2006). Many of the least-massive known exoplanets orbit M-dwarfs (Butler et al. 2004; Bonfils et al. 2005; Rivera et al. 2005), but this reflects primarily a detection bias: A planet of a given mass and orbital period will induce a larger radial-velocity variation for a lower-mass star. Whether such stars have Earth-mass planets is, of course, an open question. The habitability of such planets was recently revisited by Tarter et al. (2007), who found no compelling reasons that preclude life.

Throughout the remainder of this paper, we will quote numbers for an M4V primary (0.25 $M_{Sun}$, 0.25 $R_{Sun}$, 3200 K) and an M8V primary (0.10 $M_{Sun}$, 0.1 $R_{Sun}$, 2400 K). These two spectral types encompass that of most M-dwarfs. Our dynamics-based path favors M-dwarfs, both for their low masses and radii (as we explain below) and the fact that their low luminosity places the habitable zone much closer to the star. For the M4V primary, the equilibrium temperature of the Earth (assuming the Earth's albedo) is obtained at 0.077 AU, and for the M8V primary, it lies at 0.017 AU. A small physical separation is troublesome for imaging, but desirable for our methods. We see at least five advantages that favor the study of Earth-like planets of M-dwarfs:

1. Transits are more likely to occur. Assuming orbital planes randomly inclined to our line of sight, the probability of a transit for a planet at the orbital separations listed above is 1.5% (M4V) and 2.7% (M8V), significantly above the Earth-Sun value of 0.47%.
2. Transits are deeper and thus easier to detect. An Earth-sized planet induces transit depths of 1.3 mmag (M4V) and 8.4 mmag (M8V), as opposed to 0.084 mmag (Sun).
3. Transits are more frequent, as the orbital periods for the semi-major axes listed above are only 15 days (M4V) and 2.5 days (M8V). This is favorable for detection, since fewer hours are required to ensure sufficient orbital phase coverage. This is also favorable for spectroscopic follow-up, since there are more events per unit time and thus more total hours spent in secondary eclipse. Per year, an Earth-Sun system in an equatorial transit would spend only 13 hours in secondary eclipse, whereas the total time in eclipse is 44 hours (M4V) and 84 hours (M8V).
4. The induced stellar radial-velocity variation is much larger and commensurate with current precision. The peak-to-peak amplitude is 1.4 m/s (M4V) and 4.4 m/s (M8V), as opposed to 0.18 m/s for the reflex orbit of the Sun due to the Earth's orbit.

5. The planet-to-star contrast is much larger than that for the Earth-Sun system. In the Rayleigh-Jeans limit, this ratio depends upon the relative surface areas and brightness temperatures of the planet and star. This ratio is 0.012% (M4V) and 0.11% (M8V), compared to 0.00044% for the Earth-Sun system. This facilitates the measurement of the planetary spectrum by occultation spectroscopy.

We propose the following path for measuring the masses, radii, and emergent spectra for several Earth-like planets orbiting within the habitable zones of their stars:

I. Rocky planets transiting M-dwarfs shall be identified by transit photometry.
II. Masses shall be measured by ground-based radial velocity follow-up.
III. Infrared spectra shall be measured using the technique of occultation spectroscopy.

We explore the practical aspects of these three steps in the following sections.

**Step I. Transit Discovery.**
We advocate that the closest 10,000 M-dwarfs be surveyed photometrically with a precision and cadence sufficient to detect transits of Earth-sized planets in the habitable zone. This number is chosen so that the conclusions from a null result are of interest and commensurate with those of the *Kepler Mission*. If no such signal is found, then we would conclude that the rate of occurrence of such planets is less than 2.8% (3 sigma). However, if the rate of occurrence is only 5%, the expectation value would be 11 planets, a number worthy of the effort.

Importantly, such a survey can be carried out from the ground. Using the Mt. Hopkins 1.2m telescope with Keplercam (a thinned 4k x 4k CCD camera) in *z* band, the Transit Light Curve project (Winn et al. 2007, Holman et al. 2006) has demonstrated a relative photometric precision of 0.20 mmag (15-minute bin) for time series observations of known transiting planets (see Figure 1). Although these observers implemented modest procedures to ensure high precision (for example, the images were mildly defocused so as to increase the duty cycle and average down flat fielding errors by increasing the number of pixels in the aperture), such precision could reasonably be anticipated for similar observatories.

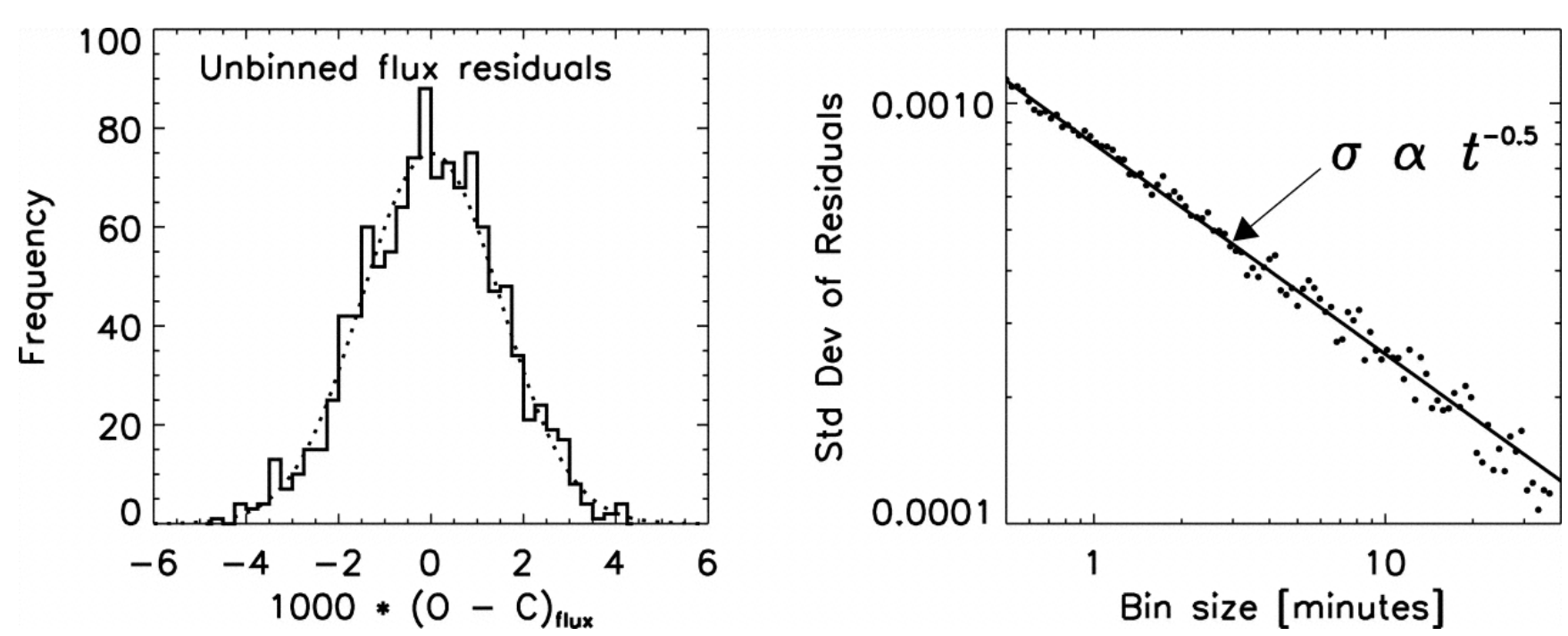

*Fig 1. (from Winn et al. 2007) Noise properties of flux residuals from time series photometry of a known transiting exoplanet system, TrES-1, gathered with the Mt. Hopkins 1.2m. Left: Distribution of unbinned residuals. The dotted line is a Gaussian function with a standard deviation of 0.15%. Right: Standard deviation of the residuals as a function of the size of the time-averaging bin size. The solid line represents the 1/sqrt(t) dependence that is expected in the absence of systematic errors.*

Unlike extant ground-based surveys that stare at fixed fields-of-view containing tens of thousands of stars, the M-dwarf targets would be spread uniformly over the sky and hence would need to be observed sequentially. Based on our experience running automated photometric observatories, we anticipate that the photometry could be reduced in real time, permitting the identification of transits in progress. This signal could then trigger a larger automated observatory either at the same site, or a more westerly longitude, which would monitor transit egress with a higher signal-to-noise. Once identified, the M-dwarf could be monitored intensively to observe a second transit and hence deduce the orbital period. In this mode, the number of hours required to survey each star is of order the orbital period, and accounting for weather losses would roughly double this requirement. This is in contrast to current transit surveys, which phase-fold archived data and typically require more than 5 times as many hours to achieve their detections (e.g. TrES-2, O'Donovan et al. 2006; HAT-P-1, Bakos et al. 2007). Astrophysical false positives (blends of eclipsing binaries that precisely mimic the desired planetary signal; O'Donovan et al. 2007) plague current wide-field surveys. This is due to the fact that the targets (1) are poorly characterized, and notably lack parallaxes, and (2) have late-F and early-G spectral types, to which the addition of an eclipsing binary with K or M spectral types yields the false positive signal. Astrophysical false positives will not be a significant source of distraction for a targeted survey of nearby M-dwarfs, since the targets will be well characterized with parallaxes, and it is extremely difficult to concoct a triple star system for which the eclipsing binary in hidden in the light of the intrinsically faint M-dwarf primary.

Although an IR survey would permit the use of smaller apertures, the relatively large cost and poor precision of infrared detectors compared to that of CCDs drives such a survey to *z*-band. Charbonneau is preparing to deploy a network of robotic telescopes that will monitor the 2000 brightest northern M-dwarfs at *z*-band with a cadence and precision sufficient to detect transits of planets with radii twice that of the Earth in habitable-zone orbits. This effort, named the MEarth Project, will likely consist of 10 14-inch telescopes located in a single roll-off enclosure and cost $0.6M. A project capable of monitoring the closest 10,000 M-dwarfs and detecting planets of Earth-radius would need to have observatories in both hemispheres and increase both the number of nodes and the typical aperture (roughly 1 m), but could nonetheless be undertaken for a relatively modest cost of order $5M. A preliminary analysis of Pan-STARRS and LSST indicates that their cadence will not be sufficient to undertake this survey.

**Step II. Mass Measurements.**

Precise radial-velocity monitoring of transiting planet candidates is required both to confirm their planetary nature and to provide precise estimates of the mass. When combined with the radius estimate, these measurements may be compared with structural models to determine the fractional composition of constituents such as iron cores, silicate mantles, and significant water envelopes (e.g. Valencia, Sasselov, & O'Connell); see Figure 2.

Transiting systems require an inclination near 90°, thus the mass of the planet may be determined directly. The peak-to-peak amplitude induced by an Earth-mass companion will be 1.4 m/s (M4V) and 4.4 m/s (M8V). Stars earlier than M5V emit sufficient flux shortward of 600nm that this signal may be measured with the well-developed optical radial-velocity techniques, either with extant facilities (such as HIRES or HARPS), of with spectrographs currently under construction (such as the New Earths Facility; see white paper by D. Sasselov). For later spectral types, or for fainter M-dwarfs of any spectral type, the spectral energy distribution requires the development of this technique at nIR wavelengths. This idea has been developed extensively in the past year in response to the call for proposals to develop the PRVS instrument for the Gemini Observatory (see white papers by H. Jones, and J. Lloyd & L. Ramsey; D. Charbonneau served as project scientist for one such proposal, named GEDI). Those efforts demonstrate that such an instrument on an 8-m observatory could achieve the requisite precision for these targets with typical integration times of 0.5 hr. Those authors address possible concerns about reduced precision due to variable telluric absorption, and the increased rotational velocities and spot coverage of stars at the bottom of the main-sequence, and find that these effects are not insurmountable. We note that the particular application of follow-up of transit-identified planets is much less time intensive that that of a radial-velocity survey. The orbital period and phase will be determined by the photometry, and the orbital eccentricity will likely be zero due to tidal circularization. Hence the only unknown Keplerian parameter will be the velocity semi-amplitude, which may be determined with only ~5 observations per star, provided the per-point precision is at least as precise as the velocity semi-amplitude.

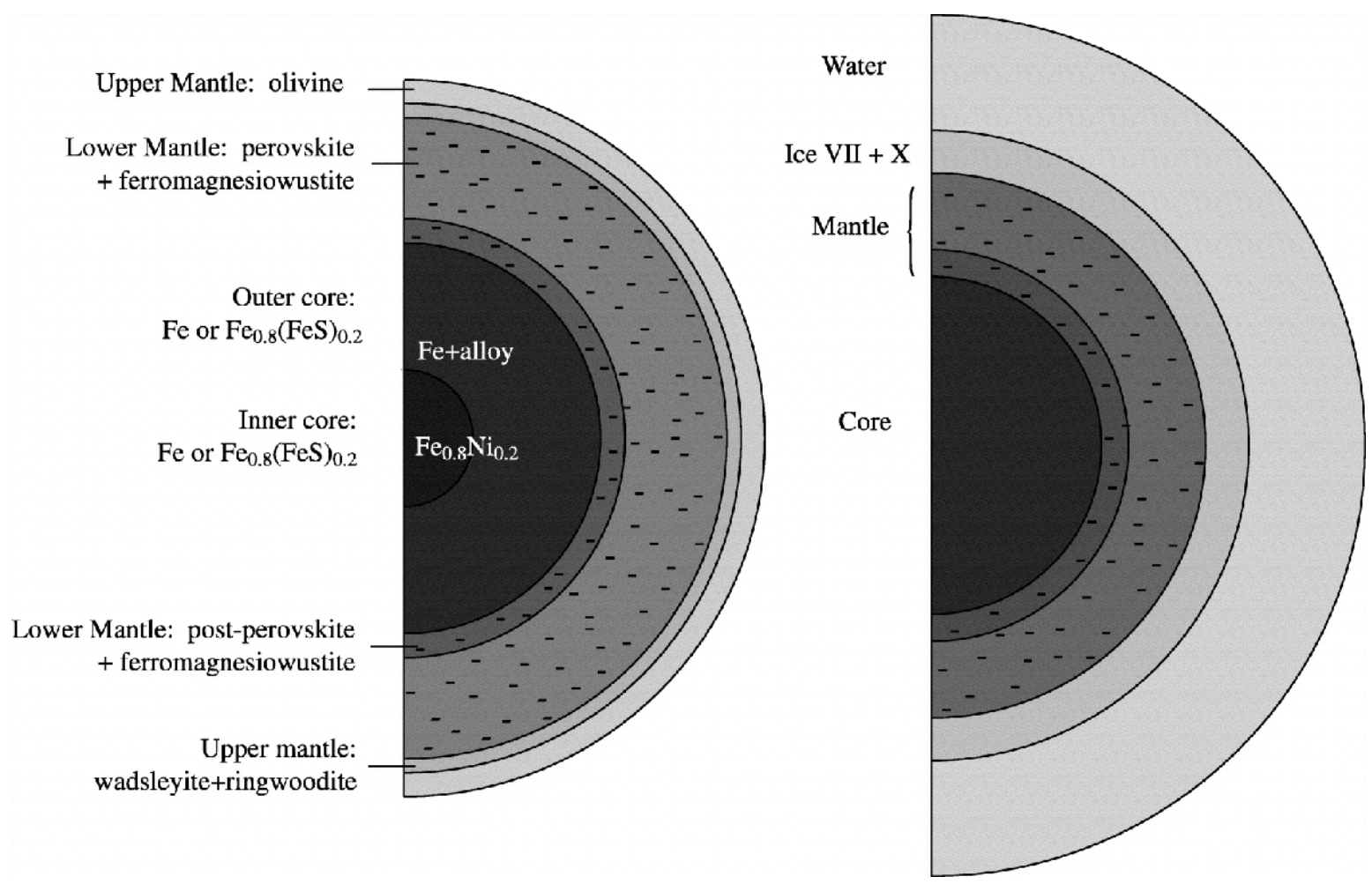

*<u>Fig 2:</u> (from Valencia et al. 2007) Schematic representation of the structural model of the least massive known exoplanet, GJ 876d. To calculate the internal structure, the authors assume a similar composition to that of Earth (left): a dense core of pure Fe or $Fe_{0.8}(FeS)_{0.2}$; a lower mantle composed of two silicate shells; and an upper mantle composed of two silicate shells. An ocean planet (right) will have an additional water/ice layer above the rocky core, and could be distinguished by its larger radius.*

**Step III. Occultation Spectroscopy.**

Measurement of the spectrum of an M-dwarf terrestrial planet is best accomplished by space-borne cryogenic IR telescopes. We do not exclude the possibility that reflected light could be detected, particularly at the longest visible wavelengths. However, the paucity of visible light emitted by cool M-dwarfs motivates IR diagnostics. The *Spitzer Space Telescope* has proven the feasibility of occultation spectroscopy and photometry for the study of giant planets orbiting close to solar-type stars. These results are described in a separate white paper (Deming et al.). Here we mention the highlights, and project the scientific return from *JWST*.

*Spitzer* investigators have detected radiation from several hot Jupiters, over six bandpasses from 3.6 to 24 μm. At the conclusion of GO-4 observations, 9 hot Jupiters will be characterized. *Spitzer* has detected day-night temperature contrasts on two hot Jupiters (Harrington et al. 2006, Knutson et al. 2007), and GO-4 observations will expand this to other planets. Spectra between 8-13 μm have been measured for two hot Jupiters (Grillmair et al. 2007, Richardson et al. 2007). The sensitivity of Spitzer to secondary eclipses measured in broad photometric bands currently extends to "hot Earths" in favorable cases. Knutson et al. (2007) detected the secondary eclipse of HD189733b at 8 μm to 60-sigma significance, and *Spitzer*'s 3-sigma radius limit for a planet in similar circumstances is just under 3 $R_{Earth}$. A GO-3 program (Deming & Seager) is currently analyzing *Spitzer* data for GJ 876 system, where the non-transiting 7.5-Earth mass inner terrestrial planet (Rivera et al. 2005) is within *Spitzer*'s detection limit at 8 μm.

Since *Spitzer* can in principle detect hot Earths, it is not surprising that "warm Earths" (300 K) are within the grasp of *JWST*. The 6.5-m cold aperture, and its thermally benign environment at L2, are reasons to anticipate success. We have calculated the S/N for occultation spectroscopy of a 290K terrestrial planet orbiting an M-dwarf, assuming that the planet emits as a blackbody in equilibrium with stellar radiation (Fig. 3). The calculation assumes parameters reasonable for *JWST*/MIRI spectroscopy. We include stellar and solar system zodiacal photon noise. Since the light from the host star will be dispersed, saturation will not occur in MIRI's minimum exposure time (3 s), and read noise will be dominated by the photon noise. We find that the S/N for spectra of the planet, at a resolving power of 100, exceed 10 for wavelengths greater than 10 μm, in a 200-hour observing program (equally divided between in-eclipse and out-of-eclipse).

*JWST* photometry of warm Earth-like planets orbiting M-dwarfs will achieve much higher S/N, capable of measuring the day-night temperature difference, just as *Spitzer* has done for close-in giant planets. A traditional concern about the habitability of planets in the habitable zone of M-dwarfs is a large temperature difference caused by tidal-locking of their rotation. Tarter et al. (2007) have argued that this is not fatal for life, in part because atmospheric circulation efficiently re-distributes heat. *JWST* photometry could determine the degree of heat redistribution for these planets, just as *Spitzer* has done for hot Jupiters. Moreover, the mere detection of a small day-night temperature difference would be compelling evidence for the existence of an atmosphere, even before spectroscopy was attempted.

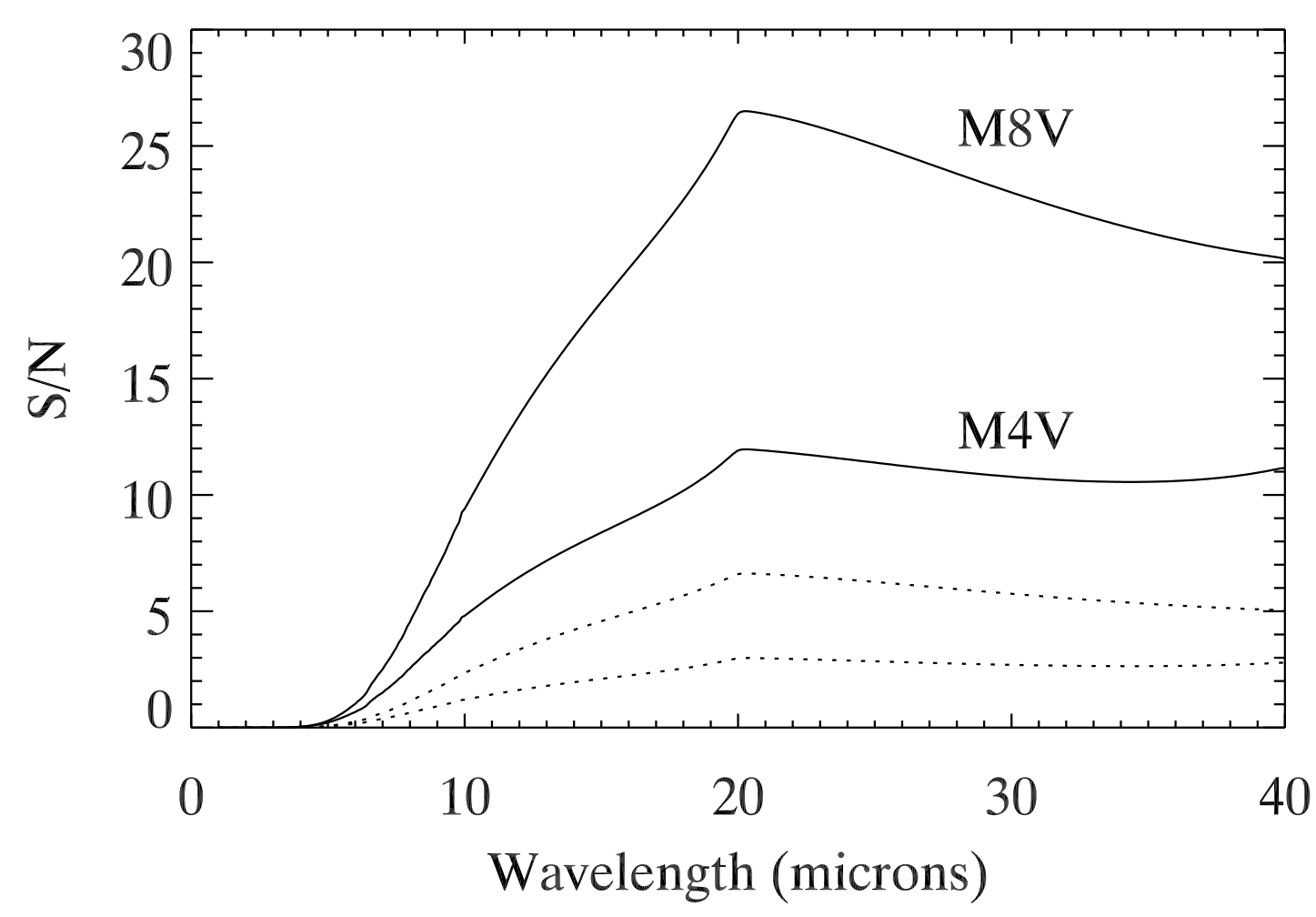

*<u>Fig 3:</u> S/N for our two cases, both having planets in the habitable zone (300 K). The spectral resolution is 100, as observed by JWST/MIRI. Planets with radii of 2 $R_{Earth}$ (solid lines) and 1 $R_{Earth}$ (dashed lines) are shown. The total observing time in these simulations for each case is 200 hours.*

## References

*"HAT-P-1b: A Large-Radius, Low-Density Exoplanet Transiting One Member of a Stellar Binary"*
G. A. Bakos et al. 2007, ApJ, 656, 552
*"Direct Detection of Exoplanets"*
J.-L. Beuzit, D. Mouillet, B. R. Oppenheimer, & J. D. Monnier 2007, in Protostars and Planets V, p. 717
"*Any Hot-Jupiter Around M Dwarfs?*"
X. Bonfils, X. Delfosse, S. Udry, T. Forveille, & D. Naef 2006, in Tenth Anniversary of 51 Peg b, p. 111
"*The HARPS search for southern planets. VI. A Neptune-mass planet around the nearby M dwarf Gl 581*"
X. Bonfils et al. 2005, A&A, 443, L15
"*Detection of Thermal Emission from an Extrasolar Planet*"
D. Charbonneau, et al. 2005, ApJ, 626, 523
*"When Extrasolar Planets Transit Their Parent Stars"*
D. Charbonneau, T. M. Brown, A. Burrows, & G. Laughlin 2007, in Protostars and Planets V, p. 701
*"Infrared Radiation from an Extrasolar Planet"*
D. Deming, S. Seager, L. J. Richardon, & J. Harrington 2005, Nature, 434, 740
"*Exploring the Frequency of Close-in Jovian Planets around M Dwarfs*"
M. Endl, et al. 2006, ApJ, 649, 436
*"A Spitzer Spectrum of the Exoplanet HD 189733b"*
C. J. Grillmair, D. Charbonneau, et al. 2007, ApJ, 658, L115
"*The Phase-Dependent Infrared Brightness of the Extrasolar Planet Upsilon Andromedae b"*
J. Harrington et al. 2006, Science, 314, 623
"*The Solar Neighborhood. XVII. 20 New Members of the RECONS 10 Parsec Sample*"
T. J. Henry et al. 2006, AJ, 132, 236
"*The Transit Light Curve Project. I. Four Consecutive Transits of the Exoplanet XO-1b*"
M. J. Holman et al. 2006, ApJ, 652, 1715
*"A Map of the Day-Night Contrast of the Extrasolar Planet HD 189733b"*
H. A. Knutson, D. Charbonneau, et al. 2007, Nature, in press
*"A Catalog of Northern Stars with Annual Proper Motions Larger than 0.15" (LSPM-NORTH Catalog)"*
S. Lepine & M. M. Shara 2005, AJ, 129, 1483
"*Nearby Stars from the LSPM-North Proper-Motion Catalog. I. Dwarfs and Giants within 33 pc of the Sun*"
S. Lepine 2005, AJ, 130, 1680
*"A Jupiter-Mass Companion to a Solar-Type Star"*
M. Mayor & D. Queloz 1995, Nature, 378, 355
"*TrES-2: The First Transiting Planet in the Kepler Field*"
F. T. O'Donovan, D. Charbonneau, et al. 2006, ApJ, 651, L61
"*Outcome of Six Candidate Transiting Planets from a TrES Field in Andromeda*"
F. T. O'Donovan, D. Charbonneau, et al. 2007, ApJ, in press, astro-ph/0610603
*"A Spectrum of an Extrasolar Planet"*
L. J. Richardson, D. Deming, K. Horning, S. Seager, & J. Harrington 2007, Nature, 445, 892
"*A 7.5 Earth-Mass Planet Orbiting the Nearby Star, GJ 876*"
E. J. Rivera et al. 2005, ApJ, 634, 625
"*The Mid-Infrared Spectrum of the Transiting Exoplanet HD 209458b*"
M. R. Swain, J. Bouwman, R. Akeson, S Lawler, C. Beichman, ApJL, submitted, astro-ph/0702593
"*A Re-appraisal of the Habitability of Planets Around M Dwarf Stars*"
J. C. Tarter et al. 2007, Astrobiology, in press, astro-ph/0609799
"*Radius and Structure Models of the First Super-Earth Planet*"
D. Valencia, D. D. Sasselov, & R. J. O'Connell, ApJ, 656, 545
*"A Decade of Radial-Velocity Discoveries in the Exoplanet Domain"*
S. Udry, D. Fischer, & D. Queloz 2007, in Protostars and Planets V, Univ. Arizona Press, p. 685
"*The Transit Light Curve Project. III. Tres Transits of TrES-1*"
J. N. Winn, M. J. Holman, & A. Roussanova 2007, ApJ, 657, 1106